# Charge Transport in Polycrystalline Graphene: Challenges and Opportunities

*Aron W. Cummings, Dinh Loc Duong, Van Luan Nguyen, Dinh Van Tuan,*
*Jani Kotakoski, Jose Eduardo Barrios Varga, Young Hee Lee\* and Stephan Roche\**

Dr. A. W. Cummings[+], J. E. B. Varga
ICN2 - Institut Català de Nanociència i Nanotecnologia, Campus UAB, 08193 Bellaterra (Barcelona), Spain.
Prof. S. Roche
ICN2 - Institut Català de Nanociència i Nanotecnologia, Campus UAB, 08193 Bellaterra (Barcelona), Spain and ICREA - Institució Catalana de Recerca i Estudis Avançats, 08010 Barcelona, Spain.
E-mail: stephan.roche@icn.cat
Dr. D. L. Duong[+], V. L. Nguyen, Prof. Y. H. Lee
IBS Center for Integrated Nanostructure Physics (CINAP), Institute for Basic Science (IBS), Sungkyunkwan University, Suwon 440-746, Korea.
Department of Energy Science, Department of Physics, Sungkyunkwan University, Suwon 440-746, Korea.
E-mail: leeyoung@skku.edu
Dr. D. V. Tuan
ICN2 - Institut Català de Nanociència i Nanotecnologia, Campus UAB, 08193 Bellaterra (Barcelona), Spain and Department of Physics, Universitat Autónoma de Barcelona, Campus UAB, 08193 Bellaterra, Spain.
Dr. J. Kotakoski
Department of Physics, University of Helsinki, P.O. Box 43, 00014 University of Helsinki, Finland and Faculty of Physics, University of Vienna, Boltzmanngasse 5, 1090 Wien, Austria
[+] The authors contributed equally to this work

Graphene has attracted significant interest both for exploring fundamental science and for a wide range of technological applications. Chemical vapor deposition (CVD) is currently the only working approach to grow graphene at wafer scale, which is required for industrial applications. Unfortunately, CVD graphene is intrinsically polycrystalline, with pristine graphene grains stitched together by disordered grain boundaries, which can be either a blessing or a curse. On the one hand, grain boundaries are expected to degrade the electrical and mechanical properties of polycrystalline graphene, rendering the material undesirable for many applications. On the other hand, they exhibit an increased chemical reactivity, suggesting their potential application to sensing or as templates for synthesis of one-dimensional materials. Therefore, it is important to gain a deeper understanding of the structure and properties of graphene grain boundaries. Here, we review experimental progress on identification and electrical and chemical characterization of graphene grain boundaries. We use numerical simulations and transport measurements to demonstrate that electrical properties and chemical modification of graphene grain boundaries are strongly correlated. This not only provides guidelines for the improvement of graphene devices, but also opens a new research area of engineering graphene grain boundaries for highly sensitive electro-biochemical devices.

## 1. Introduction

Graphene-based science and nanotechnology have been attracting considerable interest from the scientific community, in view of the numerous possibilities offered by graphene for not only studying fundamental science in two-dimensional (2D) layered structures[1,2] but also for improving the performance of flexible materials and for its integration into a variety of electrical and optical applications.[3-9] This interest is driven by graphene's superior mechanical strength and stiffness,[10] electronic and thermal conductivity,[11,12] transparency,[13] and its potential for straightforward incorporation into current silicon and plastic technologies.[14,15]

For large-area graphene, the CVD growth technique is unquestionably the best candidate for achieving a combination of high structural quality and wafer-scale growth.[16,17] Unfortunately, the transfer of graphene to diverse substrates[18,19] is still a significant challenge for a plethora of applications, including (bio)chemical sensing,[20] flexible and transparent electrodes,[16] efficient organic solar cells,[21] multifunctional carbon-based composites,[15] and spintronic devices.[22] Considerable effort is also needed for fine-tuning of the CVD growth process. In particular, the produced graphene is typically polycrystalline in nature, consisting of a patchwork of grains with various orientations and sizes, joined by grain boundaries of irregular shapes.[23,24] The boundaries consist of an approximately one-dimensional (1D) distribution of non-hexagonal rings,[23,24] and appear as structural defects acting as a source of intrinsic carrier scattering, which limits the carrier mobility of wafer-scale graphene materials.[25]

Graphene grain boundaries (GGBs) also introduce enhanced chemical reactivity.[26] This opens a hitherto unexplored area of research, namely, GGB engineering of the properties of polycrystalline graphene, with further diversification of material performance and functionality. Selective chemical functionalization of GGBs with various functional groups and selective adsorption of various metal particles not only modify the carrier mobility of polycrystalline graphene but also make it biochemically active, a feature which could be utilized in highly sensitive biochemical sensors. With the capability of engineering GGBs during CVD growth and their applications mentioned above, a new multidisciplinary field of science and engineering can be established. Although grapheen oxide is another category of graphene with strong chemical functionalization, the materials exist in a powder form and their use is also different from large area CVD-grown graphene. The extensive review on this has been published elsewhere.[27-32] We limit our discussion to large-area CVD-grown polycrystalline graphene here. In this review, we present the current progress of this field through an overview of the experimental efforts to understand the fundamental connection between the structure and the corresponding mechanical, electrical, and chemical properties of polycrystalline graphene. We also show why nanotechnology and related methods are essential not only for observing and analyzing GGBs, but also for tailoring nanomaterials with superior performance.

## 2. Structure and Morphology of GGBs

### 2.1. GGBs formed between two domains with different orientations

While a detailed description of graphene defects has been extensively reviewed already,[33-36] here we point out and update some important features of GGB structures. This will aid in understanding the physical and chemical properties of GGBs, with an aim toward controlling their behavior and functionality. GGBs are formed at the stitching region between two graphene domains with different orientations or with a spatial lattice mismatch. In general, a GGB is a thin meandering line that consists of a series of pentagonal, hexagonal, and heptagonal rings,[23-25] where the structure and periodicity of the GGB are determined by the misorientation angle between two domains. An example of this is shown in the top panel of **Figure 1a**, which depicts a 5-7 GGB formed between two grains with a misorientation angle of 21.8°. This GGB consists of a periodic series of pentagon-heptagon pairs. In comparison, the bottom panel of **Figure 1a** shows a high-resolution transmission electron microscopy (TEM) image of a GGB between two domains with a misorientation angle of 27°. While the experimental image indicates a non-straight GGB, it also consists of a single thin line of pentagon-heptagon pairs.[23]

However, this simple GGB structure is not always achieved during the CVD growth process. For example, **Figure 1b** shows a theoretical model (left panel) and observation by scanning tunneling microscopy (STM; right panel) of a disordered GGB consisting of a complex and meandering series of various carbon rings, as well as the occasional vacancy defect.[37,38] In this type of structure, the electronic effect of the GGB can extend to several nanometers in width, as can be directly observed from the STM image. Its corresponding transport properties are independent of the orientation of the two domains forming the GGB.[37] In order to minimize the structural energy due to the presence of non-hexagonal rings, the GGB and the surrounding graphene grains can lead to buckling along the length of the GGB.[39,40] This is true even in the ideal case, and thus is a common feature of all GGBs. For example, the top panel of **Figure 1c** shows the morphology of a three-dimensional (3D) model of a GGB and its neighboring grains, indicating that out-of-plane buckling can occur.

The bottom panel of **Figure 1c** shows buckled graphene morphology on copper measured before and after ultraviolet (UV) treatment.[40] The buckling line at position 2 coincides with the buckled GGB visualized after UV exposure.

The existence of GGBs can strongly alter the mechanical properties of polycrystalline graphene. While monocrystalline graphene has been established as the strongest material ever measured, with an intrinsic strength of 42 N m$^{-1}$, a failure strain of 0.25, and a Young's modulus of 1 TPa,[10] the mechanical properties of polycrystalline graphene remain under intense scrutiny. The usual method for estimating the elastic properties of 2D materials is to transfer the membrane onto a substrate with an array of holes, and apply a force to the membrane through one of the holes with an atomic force microscope (AFM).[10] The first reported measurements indicate that GGBs in CVD-grown graphene significantly lower the elastic constant by a factor of six,[23,41,42] with an average breaking load of about 120 nN, an order of magnitude lower than for monocrystalline graphene.[10] The strength of individual GGBs was also found, theoretically and experimentally, to strongly depend on the misorientation angle between graphene domains.[43-47] However, these results are for a single GGB between two domains, and it is uncertain how they translate to macroscopic samples containing several GGBs. Moreover, the cracks that appear upon failure do not necessarily follow the GGBs but can also penetrate through the grains,[40,48] even if they originate at the GGB regions.

A more realistic model for polycrystalline graphene can be constructed by simulating seeded growth of separated graphene grains with random orientations, and allowing such grains to merge together to form natural GGBs.[37] For these samples, the angle-dependence of the mechanical properties vanishes, and clear trends appear as a function of the average grain size. Increasing grain sizes lead to lowering fracture strain and increasing elastic modulus, whereas the variation in the strength of the material is much less affected, being about 50% of that of monocrystalline graphene.[37] The cracks originate at GGB junctions, and propagate through the grains, in agreement with the experiments.[48] More restricted models containing several connected hexagonal graphene grains have recently confirmed these findings.[49]

Although much progress has been made in understanding the mechanical strength of polycrystalline graphene, questions still remain. For example, the breaking loads for early measurements[23,42] differ significantly from those measured more recently.[46,47] In addition, as noted above, the applicability of the AFM measurements to macroscopic samples remains an open question. To finally resolve the issue, we would need a new measurement technique for estimating the elastic properties of 2D materials, which would avoid the shortcomings of the method utilizing an AFM tip.

**2.2. GGBs formed between two domains with the same orientation**

In addition to degraded mechanical properties, numerous studies have shown that carrier transport in polycrystalline graphene is strongly affected by GGBs.[40,50-53] Therefore, a great deal of effort has been made to eliminate the formation of GGBs during CVD by growing monocrystalline graphene.[54-59] There are two primary methods to obtain monocrystalline graphene with CVD. One method is to control the number of nucleation seeds (and thus the individual grain size) by polishing the copper substrate,[59] annealing it at high temperature before growth,[54,55] or using copper oxide.[56,57] Recently, this approach has been able to realize CVD growth of individual grains on the order of several millimeters in diameter. The drawback of this method is that it takes a long time (for instance 12 hours) for a single graphene grain to grow to a large size. Furthermore, the crystallinity within a single domain is not guaranteed or at least not confirmed rigorously. Another method is based on controlling the orientation of graphene domains, such that their crystal lattices are aligned.[60-63] One would then expect that these domains will merge cleanly, without forming any GGBs at the stitching regions, as shown in the upper left panel of **Figure 1d**. However, experiments have shown that this is not always the case. For example, no GGBs were found in the case of graphene growth on a monocrystalline boron nitride (BN) flake[64] (red circle, top right panel of **Figure 1d**). On the other hand, a line of 5-8-5 rings was observed for graphene grown on nickel (Ni; bottom panels of **Figure 1d**) even though the graphene domains have the same orientation.[65] This is caused by a translational mismatch between neighboring grains. In addition, non-straight edges can also lead to more complex GGB structures than the 5-8-5 example shown here. Therefore, additional proof such as high resolution STM, TEM, or electrical transport measurements are necessary to confirm the absence of GGBs in these samples. Different methods of observing GGBs are described below.

## 3. Methods of Observing GGBs

To study the properties and structure of GGBs, or to control the graphene growth process, it is necessary to develop methods to determine the location of the GGBs. This information is not straightforward to obtain due to the atomic width of the GGBs (on the nm scale), and is even more challenging for large-scale observations. A primitive approach is to stop the CVD process before graphene growth is complete. Then, the GGB location can be roughly estimated as the stitching region between two domains.[59] However, graphene domains are not typically monocrystalline and thus a large number of GGBs can be missed with this approach.[40,50,59]

Another approach to determine the location of the GGBs relies on mapping the orientation of the graphene grains; the shape of each grain is identified, and the GGB locations are then indirectly determined at their boundaries. The techniques for determining the grain orientation include TEM,[23,24] low electron energy microscopy,[63] and polarized optical microscopy (POM) of spin-coated liquid crystals on graphene.[66-68] However, these methods will not reveal boundaries between grains with the same orientation. An alternative method, which sidesteps this problem, is to directly observe the location of the GGBs by taking advantage of their chemical properties.[34] These methods are discussed in more detail below.

### 3.1. TEM

The principle of using TEM to map the graphene grain orientations is shown in **Figure 2**.[23] The diffraction pattern of monocrystalline graphene is six-fold symmetric, corresponding to the symmetry of the honeycomb lattice. If the observed region includes two different orientations, the diffraction pattern consists of two different hexagons rotated by a specific angle, as shown inof **Figure 2a**. This is the misorientation angle between the two grains. By doing this analysis over the entire sample, one can map the orientation of the graphene lattice at each point in the sample. An example is shown in **Figure 2b**, where the colored regions mark grains of different orientations.

### 3.2. Liquid crystal deposition

Although TEM observations provide atomic resolution of GGBs at a nanometer scale, the GGB distribution at millimeter or centimeter scales is not easily accessible. Here, we describe several methods of observing GGBs at large scale. **Figure 3a** shows the principle of using liquid crystal (LC) (4-Cyano-4'-pentylbiphenyl; 5CB) molecules to observe graphene grain orientation with POM.[67] A 5CB molecule consists of two hexagonal benzene rings with a nitrogen atom at one end and a long carbon chain at the other end. It is expected that the hexagonal rings of the 5CB molecule will align along the graphene lattice with AB stacking order. Graphene grains with different orientations provoke the 5CB molecules to align in different directions depending on the grain orientation, which can be observed as a contrast difference using POM. This can be seen in **Figure 3b**, which shows two POM micrographs that indicate a clear contrast between graphene grains of different orientations. This approach can be extended to a large scale, as shown in the right panel of **Figure 3b**.

Interestingly, experiments have not revealed a three-fold symmetry for the alignment of the 5CB molecules on graphene, which would be theoretically expected. Further studies are required to fully understand the rearrangement of LC molecules.

### 3.3. UV treatment

Instead of mapping the orientation of each graphene grain, the high chemical reactivity of GGBs can be utilized for their direct visualization.[40,69,70] One approach involves the use of an oxidizing agent to selectively oxidize copper underneath the GGBs.[40] **Figure 4a** shows the principle of UV treatment of graphene on a copper substrate in a humid environment. O and OH radicals are generated under UV exposure, and these radicals can easily invoke strong chemical reactions near the defect sites. In particular, GGBs, aggregates of defects such as vacancies, pentagons and heptagons, are most vulnerable for radical attack. These radicals penetrate through graphene defects at the GGBs to oxidize the underlying copper substrate, forming copper oxides. This provokes volume expansion to several hundred nm in the region of the GGB lines, and these oxidized lines can then be observed under an optical microscope. **Figure 4b,c** are optical and AFM images of the graphene sample after UV treatment, clearly indicating the positions of the GGBs.

It is worth noting that the methods discussed in this section are complementary to each other, where a combination of techniques can be used to visualize GGBs from the atomic scale to the wafer scale. LC coating and overlapping two graphene layers can easily determine the location of GGBs when the grains have different orientations. However, it is not possible to use these methods to determine if two grains have the same orientation. In this case, TEM, STM, or the UV oxidation methods are required.

### 4. Measurement of Electrical Transport across GGBs

In addition to their structural characterization and identification, it is important to understand how the GGBs influence electrical transport phenomena in polycrystalline graphene. Various measurements have been made to understand the electrical properties of GGBs. These measurements fall into three primary approaches. The first approach involves local two-point measurements, which are accomplished with STM and scanning tunneling spectroscopy (STS).[38,71-75] With these measurements, it is possible to deduce the local electronic density of states, the local charge density, and the charge scattering mechanisms associated with GGBs, thus permitting the spatially-dependent electrical characterization of GGBs at the atomic scale. The second approach involves four-probe measurements, which can be used to analyze the influence of individual GGBs at a scale of several micrometers.[50,51,53] By subtracting the contribution of each graphene grain from an inter-grain resistance measurement, the resistivity of a single GGB can be estimated. In combination with microscopic or spectroscopic techniques, this approach allows one to correlate the resistivity of a single GGB with its structural or chemical properties. Finally, the global impact of GGBs can be studied by measuring the sheet resistance of polycrystalline graphene samples over a wide range of average grain sizes and distributions, which are tunable by the CVD growth conditions. By employing a simple scaling law (as discussed below), it is then possible to extrapolate the average GB resistivity.[40,76] Taken together, these measurement techniques provide the electrical characterization of GGBs at various length scales, thus helping to reveal a comprehensive picture of charge transport in polycrystalline graphene. A more detailed overview of these methods is given below.

### 4.1. Two-probe measurements

Two-probe STM and STS techniques can be used to locally study the electrical properties of GGBs.[38,71,73-75] By varying the voltage and position of the STM tip, it is possible to determine the nature of localized states, the charge doping, and the local scattering mechanism corresponding to a given morphology of the GGB. One example of such analysis is shown in **Figure 5a-b**.[75] **Figure 5a** shows the differential tunneling conductance, dI/dV, taken at various points on (blue curves) and next to (red curves) a GGB in CVD-grown graphene. A STM profile of the GGB and the points where the measurements were made is shown in **Figure 5b**. These results indicate the presence of a peak in the tunneling conductance near the Dirac point whenever the STM tip lies on top of the GGB. Meanwhile, this peak does not appear for measurements away from the GGBs. Density functional theory (DFT) calculations have attributed this peak to the localized states arising from two-coordinated carbon atoms in the GGBs.[75] The STM map (not shown here) also reveals interference superstructures due to scattering from the GGBs, indicating the contribution of significant inter-valley scattering. This supports the hypothesis about the presence of two-coordinated atoms, since inter-valley scattering stems from atomic-scale lattice defects.[77]

**Figure 5c** shows another map of dI/dV curves as the STM tip is scanned across a GGB.[65] Similar to **Figure 5a**, an enhanced local density of states is observed at positive voltage when the tip is located over the GGB. The voltage associated with the minimum of dI/dV, as shown in **Figure 5d**, indicates a strong negative shift around the position of the GGB, revealing n-type doping of the GGB compared to bulk p-type doping of the graphene grains. This shift in doping corresponds to an electrostatic potential barrier of a few tens of meV. Finally, STM interference patterns indicate that some GGBs are dominated by inter-valley scattering while others are dominated by backscattering. The type of scattering appears to depend on the structure of the GGB, where a GGB consisting of a continuous line of defects shows primarily backscattering behavior and a periodic line of isolated defects is dominated by inter-valley scattering.

Other STM studies of GGBs reveal similar results to those mentioned above, with GGBs forming p-n-p or p-p'-p junctions with the bulk-like graphene grains, where p' < p. The doped regions associated with the GGBs are on the order of a few nm wide, showing an abrupt transition between the GGB and the grain.[38,74] Other works reveal the presence of localized states along GBs in graphene and graphite.[65,71,78] In general, STM/STS studies indicate that GGBs are a source of localized states and electrostatic potential barriers in polycrystalline graphene, and can serve as significant sources of charge scattering.

**4.2. Four-probe measurements**

In order to make a four-probe measurement of the resistivity of a GGB, it is necessary to first identify its location. This can be done, e.g., with non-destructive TEM or by drop-casting a liquid crystal layer.[50,67,68] In the case of two regular hexagonal graphene domains merged together, simple optical microscopy can also be used to identify the boundary location, as shown in the grey background of **Figure 6**. A Hall bar is then fabricated by e-beam lithography, and a regular four-probe measurement is performed to determine the resistance of the left (L) domain, the right (R) domain, and the middle (M) region between the two domains. A constant current is applied from the left to the right while the voltage drop between two adjacent electrodes is measured, and the resistance is calculated by Ohm's law, $R_L = V_L/I$, $R_R = V_R/I$, and $R_M = V_M/I$. In general,

$$R_M = mR_L + R_B + nR_R = \alpha R_D + R_B,$$

where $m+n=\alpha$ (due to the $\alpha L$ length of the middle part) and $R_D$ is average resistance of the graphene domains, $R_D = \dfrac{mR_L + nR_R}{m+n}$. If the samples are uniform $(R_L = R_R = R_D)$ or if the GGB is located precisely in the middle ($m = n$), then the resistance of the GGB is determined. Otherwise, the precise location of the GGB needs to be determined to extract its resistance. The resistivity of the GGB $(\rho_{GB})$ is calculated from[50]

$$R_M = \alpha R_D + \frac{\rho_{GB}}{W}.$$

Note that $\rho_{GB}$ has the same dimensions as bulk resistivity ($\Omega-m$). The relationship between $\rho_{GB}$ and bulk resistivity $\left(\rho_{GB}^{bulk}\right)$ is

$$R_B = \frac{\rho_{GB}}{W} = \frac{\rho_{GB}^{bulk} \cdot l_{GB}}{t \cdot W},$$

where $l_{GB}$ and $t$ are the effective width and thickness, respectively, of the GGB.

As described above, four-probe measurements are a useful tool for addressing the electrical transport properties of individual GGBs. With this measurement technique, the contribution of within the grains can be separated from the inter-grain resistance, and by normalizing for the length of the GGB, the characteristic transverse GGB resistivity $\rho_{GB}$ is derived. These measurements also yield useful information about the performance of devices based on CVD graphene, because the measurements are made in a device configuration. An example of the experimental setup and measurement results can be seen in **Figure 7a-b**.[51,53] **Figure 7a** is an optical image of the four-probe measurement setup across an approximately 4-μm-long GGB. **Figure 7b** shows the I-V curves corresponding to the left and right grains (red and blue curves) and across the GGB (green curve). Here, the I-V curves indicate a much larger inter-grain resistance compared to the resistance measured within each grain, indicating extra scattering provided by the presence of the GGB. This particular measurement yielded a GB resistance of 2.1 kΩ, or $\rho_{GB}$ = 8 kΩ.μm when scaled by the GGB length. Temperature-dependent measurements show that $\rho_{GB}$ is insensitive to temperature, pointing to a defect-induced scattering mechanism. Magnetotransport measurements reveal the presence of weak localization at low temperatures,[51,53] indicating that GGBs are significant sources of inter-valley scattering, in agreement with the STM studies mentioned above.

A similar measurement setup is shown in **Figure 7c**, on a device fabricated on a specially prepared TEM window that allows for concurrent transport measurements and identification of the individual grains and the GGB.[50] An example of the measurement results can be seen in **Figure 7d**. In the top graph, the gray curves correspond to the resistance measured within each grain, while the black curve is the inter-grain resistance. In the bottom graph, the green curve shows the extracted GB resistivity as a function of applied gate voltage. Here, $\rho_{GB}$ peaks at a value of 4 kΩ.µm at the Dirac point. With the four-probe measurements, $\rho_{GB}$ has been extracted for CVD graphene prepared under several growth conditions, and it has been shown that the resistivity depends strongly on the structure of the GGB. For example, a growth procedure yielding well-connected grains gives $\rho_{GB}$ = 1 to 4 kΩ.µm at the Dirac point, while a growth procedure yielding poorly-stitched grains results in values of $\rho_{GB}$ an order of magnitude larger. Interestingly, some overlapping GBs have a negative resistivity, with the inter-grain resistance smaller than the combined resistance of the individual grains. This is attributed to reduced scattering in the double-layer overlapped region compared to the single-layer grains.

### 4.3. Global measurements from scaling law

In general, GGBs are formed randomly during the CVD growth process, and their electrical properties are not uniform. Therefore, in addition to studies of individual GGBs, it is also necessary to study GGBs on a large scale to extract a reliable average of their transport properties. This average quantity is represented by the GB resistivity $\rho_{GB}$, which can be extracted from an Ohmic scaling law, as illustrated in **Figure 8**. **Figure 8a** shows a 1D model of $n$ graphene grains separated by $n$ GBs. The sample resistance $R$ includes the resistance of the $n$ grains $R_i^G$, and the resistance of the $n$ GGBs $R_i^{GB}$ ($R = \sum_{i=1}^{n} R_i^G + \sum_{i=1}^{n} R_i^{GB}$). These terms can be written as $R = R_S \cdot L/W$, $R_i^G = R_{S,i}^G \cdot L_i/W$, and $R_i^{GB} = \rho_i^{GB}/W$, where $R_S$ is the overall sample sheet resistance, $R_{S,i}^G$ is the sheet resistance of each grain, $\rho_i^{GB}$ is the resistivity of each GB, and $L_i$ is the length of each grain. Putting all this together, the sample sheet resistance can be written as $R_S = \sum_{i=1}^{n} R_{S,i}^{G} \cdot \frac{L_i}{L} + \sum_{i=1}^{n} \frac{\rho_i^{GB}}{L}$. The first term is the average sheet resistance of the graphene grains $R_S^G$, which is independent of $n$, while the last term strongly relies on $n$ or the grain size. This term is equivalent to $n \cdot \rho_{GB}/L = \rho_{GB}/l_G$, where $\rho_{GB}$ is the average GB resistivity and $l_G$ is the average grain diameter. The final expression is $R_S = R_S^G + \rho_{GB}/l_G$, where $R_S$ can be measured by the Van der Pauw method, as shown in **Figure 8b-c**. The average grain size can be estimated by visualizing the GB structure of the sample or with Raman measurements, as described in the main text. By measuring the sheet resistance of samples that span a range of average grain sizes, one can extract $R_S^G$ and $\rho_{GB}$, as shown in **Figure 8d**.

The two- and four-probe measurement techniques yield valuable information about the electrical properties of GGBs at the atomic and individual-grain scales. These microscopic electrical properties can be correlated to the macroscopic ones, which are applicable to the analysis of experimentally available large-area graphene. This can be accomplished with the global scaling law, as discussed above. Two examples of this procedure are given in **Figure 9a-b**. **Figure 9a** shows a series of sheet resistance measurements over several orders of magnitude of average grain size.[76,77,79-82] Applying the scaling law to this data (black line in **Figure 9a**) results in $\rho_{GB}$ = 0.67 kΩ.μm. This value is somewhat lower than those obtained in the four-probe measurements mentioned above. However, because the measurements did not involve back gate modulation, it is likely that the sheet resistance was measured away from the Dirac point, resulting in a lower value of $\rho_{GB}$. It should also be noted that the x-axis of **Figure 9a** was obtained through the D/G ratio in Raman spectroscopy, and thus represents an average distance between defects rather than the true grain size.

Another example of the scaling behavior is shown in **Figure 9b**.[40] In this case, the grain sizes are estimated with an optical microscope, and a fit to the scaling law gives $R_S^G$ =

130 Ω and $\rho_{GB}$ = 1.4 kΩ.μm. One useful consequence of using the scaling law is that it allows for an estimate of the average sheet resistance within the grains, $R_S^G$ (for a good fit, it is best to have a range of grain sizes such that $R_S^G < \rho_{GB}/l_G$ for the smallest grains and $R_S^G > \rho_{GB}/l_G$ for the largest grains). For example, based on the extracted values of $R_S^G$ and $\rho_{GB}$, the GGBs begin to dominate the sheet resistance of these samples when the average grain size is less than $l_G = \rho_{GB}/R_S^G \approx 10\,\mu m$. This information can serve as a useful design parameter when considering large-scale applications of polycrystalline graphene.

## 5. Manipulation of GGBs with functional groups

### 5.1. Chemical reactivity of GGBs

In addition to the general electrical transport properties of polycrystalline graphene, the chemical properties (reactivity, functionalization, etc.) of GGBs have been extensively discussed. For example, it has been shown theoretically that non-hexagonal atomic arrangements in graphene, such as the Stone-Wales defect, yield higher chemical reactivity than the ideal hexagonal structure,[83-87] and this behavior has been extended to GGBs. A schematic representation is shown in **Figure 10a**, where oxygen atoms preferentially attach to the non-hexagonal sites located in the GGBs. Selective oxidation of GGBs can be demonstrated by transferring CVD graphene to a mica substrate and heating the sample for 30 minutes at 500°C. This process selectively burns away the GGBs,[69] giving access to the grain morphology within the samples with AFM. A representative AFM image is given in **Figure 10b**, where the dark lines indicate the location of the removed GGBs. This procedure not only provides a simple means of characterizing the grain morphology in the samples but also highlights the enhanced chemical reactivity of the GGBs.

UV treatment of polycrystalline graphene on a copper substrate can also reveal selective functionalization of the GGBs.[40] Under humid environment, O and OH radicals generated by the UV light preferentially attach to the GGBs, making the defects at the GGBs inert. This allows next incoming radicals to diffuse through large-pore heptagons and higher-order defects to eventually oxidize and expand the underlying copper substrate, as explained above. The degree of volume expansion can be engineered by controlling oxidation times, and the morphological changes around GGBs are easily identified by AFM and optical microscopy. The dark lines in **Figure 10c** reveal the grain structure of the polycrystalline graphene. The grain structure is also revealed via Raman mapping of the sample, as shown in **Figure 10d-g**. **Figure 10d** outlines the formation of a strong D-band associated with the GGBs after UV treatment. The D-peak also forms within the graphene grains, but its magnitude is much smaller, highlighting the higher chemical reactivity of the GGBs. Redshifts of the G and 2D (G') bands in the GGBs after UV treatment are attributed to strain induced by the oxidized copper below the GGBs. **Figure 10e-g** show that after UV treatment, spatial mappings of the D, G, and 2D peaks correlate well with the optical image of the GGBs. It should be noted that Raman mapping shows no evidence of the GGBs prior to UV treatment, indicating the strong influence by the oxidation of the GGBs.

The experimental demonstrations of the chemical reactivity of GGBs reported to date suggest that polycrystalline graphene may be a good material for the development of chemical sensors. For example, gas sensors based on pristine (single-grain) and polycrystalline graphene have yielded highly different responses to toluene and 1,2-dichlorobenzene, with the polycrystalline graphene sensor showing a response 50x greater than that of pristine graphene.[26] This improvement in the sensitivity of the sensor is attributed to the increased reactivity of the GGBs and the enhanced impact that line defects have over point defects on transport features in two dimensions. This highlights the combined role that chemistry and charge transport play in the electrical properties of polycrystalline graphene.

**5.2. Selective functionalization of GGBs**

As described above, GGBs are more chemically active than the graphene basal plane. However, selective functionalization of GGBs with an appropriate reactant is still an on-going area of research. Our main concern is a selective functionalization of GGBs, although defects inside grain could be functionalized as well. The whole graphene layer still retains metallicity with slightly increased sheet resistance. This is good contrast with heavily functionalized graphene oxide that leads to an insulator. Ozone is a good candidate for this purpose because it is inert with the graphene basal plane.[88,89] **Figure 11, 12** shows measurements of the electrical reponse of the graphene basal plane and GGBs to ozone generated by UV exposure under an $O_2$ environment. A four-probe device was fabricated on the merged region of two graphene domains (described in **Figure 6**), as shown in **Figure 11**. Series of Hall bar geometry (5x5 $\mu m^2$) was fabricated across through an expected GGB line as shown in **Figure 11a**. The final device is shown in **Figure 11b-c** after graphene parterning, metal depostion and lift-off process. by e-beam lithography. After fabrication processes including graphene transfer and e-beam lithography, the GGBs and partial graphene basal plane are expected to be contaminated. Therefore, the sample was heat-treated at different conditions under vacuum ($10^{-2}$ Torr). Physical adsorbates were simply removed at 150 $^o$C for one hour, and the transport characteristics of the grains and the GGB were measured, as shown in **Figure 12a**. Here, the black and blue lines represent the intra-grain resistances $R_L$ and $R_R$, and the red line is the merging region resistance $R_B$. As expected, $R_B$ is larger than $R_L$ and $R_R$, due to the extra resistance contributed by the GGB. Next, the sample was further annealed at 250 $^o$C for 3 hours. **Figure 12b** shows that the resistance of the graphene basal plane was not changed, while the resistance across the GGB decreased significantly. This decrease in resistance implies that functional groups at the GGB were removed, as supported by the simulation results in the next section. The sample was then exposed to UV under an $O_2$ enviroment (0.5 Torr). The resistance across the GGB increased, while the resistance of the graphene basal plane was still unchanged, as shown in **Figure 12c**. This strongly suggests that the GGBs are selectively functionalized by ozone generated by UV. This systematic series of measurements leads us to conclude that the GGBs can be selectively functionalized by ozone. This is a key step towards further biochemical modification of GGBs. We notice that the UV treatment is saturated after 1 minute exposure. Longer time UV exposure doesn't increase the resistance at GGBs.

### 5.3. Effect of functional groups on electrical transport at GGBs by simulation

As discussed above, the resistance at the GGBs can be modified by changing their functional groups. Proving this concept with a current measurement technique is a challenge because the chemical reaction occurs on the nanometer scale at the GGBs. Therefore, numerical simulation is a key strategy to understand this process. Several theoretical and numerical approaches have been employed to study charge transport across individual GGBs.[25,52,90-93] Here, an approach which allows the study of large-area polycrystalline graphene with a random distribution of GB orientations and morphologies is outlined. The polycrystalline graphene sample is created using molecular dynamics simulations that mimic the growth of CVD graphene,[37] and its electrical properties are described with the tight-binding formalism. An example of a small portion of a polycrystalline sample is shown in **Figure 13a**. To study transport, the time evolution of an electronic wave packet within the graphene sample is tracked.[94] The conductivity can then be calculated with the Kubo formula $\sigma(E,t) = \frac{e^2}{4}\rho(E)\frac{\partial}{\partial t}\Delta X^2(E,t)$, where $\rho(E)$ is the DOS and $\Delta X^2(E,t)$ is the mean-square spreading of the wave packet. **Figure 13b-d** shows some snapshots of the time evolution of a wave packet within a polycrystalline graphene sample, highlighting the scattering and localizing effects around the GGBs. By assuming a wave packet that initially covers the entire sample, one can get a global picture of the scattering induced by GGBs. Once the conductivity is known, the sheet resistance is given by $R_s = 1/\sigma$. By doing this simulation for a range of average grain sizes, the GGB resistivity can be extracted using the scaling law described in section 4.3. To include the effect of chemical functionalization, adsorbates are randomly attached to the GB atoms at different concentrations (as illustrated in **Figure 10a**). Tight-binding parameters for describing hydrogen, hydroxyl, and epoxy groups have been taken from the literature.[95-97]

**Figure 14a-b** shows a typical example of a 5-7 GGB functionalized by O and OH groups, respectively. The resistivity of the GGBs with different functional groups at various concentrations is extracted, as shown in **Figure 14c**, where $\rho_{GB}$ is plotted as a function of adsorbate coverage, defined as the number of adsorbates relative to the total number of GGB atoms in the sample. For coverage greater than 100%, the adsorbates are allowed to functionalize the carbon atoms next to the GGBs. For all types of adsorbates, $\rho_{GB}$ increases with coverage, regardless of their type. However, it is also noted that $\rho_{GB}$ is strongly adsorbate-dependent. For example, while both H and OH groups are chemisorbed to the top site of a single carbon atom, H groups have a stronger effect on transport through the GGBs than OH groups, with $\rho_{GB}$ nearly 4 times larger at 200% coverage. This difference can be ascribed to the electronic structure of each type of adsorbate. The simulations employ a resonant scattering model, where each adsorbate is characterized by an on-site energy $\varepsilon_{ads}$ and a coupling to a single carbon atom $\gamma_{ads}$. The net effect of this model is to introduce an energy-dependent scattering potential,[95] $V_{ads}(E) = \gamma_{ads}^2/(E - \varepsilon_{ads})$. Using parameters for H and OH taken from the literature,[94,95] this gives $V_H(E=0) = -40\gamma_0$ and $V_{OH}(E=0) = 1.8\gamma_0$. Since $\sigma_{DC}$, and hence $R_S$ and $\rho_{GB}$, are calculated at the Dirac point, the H groups present a much stronger scattering potential than the OH groups. Calculations have also shown that H groups induce strongly localized states near the Dirac point, while OH adsorbates result in a more dispersive impurity band lying in the valence band of graphene.[95] Meanwhile, the O group chemisorbs in the bridge site by forming a pair with adjacent carbon atoms in the graphene

lattice (epoxide).[97] The simulations clearly show that the resistance at GGBs with functional groups is much higher than that of pure GGBs. **Figure 14d** shows a summary of the values of $\rho_{GB}$ derived from measurements compared to the simulation results.[40,50,51,53,76,98] The solid symbols are from the electrical measurements described earlier in this review, and the open symbols are the numerical simulations. Here, most measurements give $\rho_{GB}$ in the range of 1 to 10 kΩ.μm, except for one that gives values one to two orders of magnitude smaller.[98] This difference could be caused by the measurement technique, where $\rho_{GB}$ was measured with four-probe STM under ultra-high vacuum, while the other groups fabricated physical contacts on their samples. This extra fabrication step could lead to additional contamination, increasing $\rho_{GB}$. Accordingly, the numerical simulations show that it is possible to bridge the gap between the various measurements by systematically increasing the amount of chemical functionalization of the GGBs. The situation becomes more complicated by several other parameters such as the structure and resistivity of the GGBs, as mentioned previously.[52] This is highlighted by the measurements labeled "small grain" and "large grain" in **Figure 14d**, where growth conditions yielding large grain samples also tend to yield poorly connected and highly resistive GGBs.[50] Nevertheless, these results highlight the strong impact that chemical functionalization can have on the electrical properties of GGBs.

## 6. Challenges and Opportunities

The observation and characterization of GGBs at both atomic and macroscopic scale is mandatory to understand the transport properties and the related underlying physics and chemistry of polycrystalline graphene. As described in our review, TEM and STM, combined with theory and simulation, can provide information at the atomic scale, with the related transport properties revealed with the assistance of STS. UV-treatment and liquid crystal coating, combined with optical microscopy, can provide information on both the grain boundary distribution at the macro scale and the orientation of each domain, while macroscopic transport properties can be derived using the scaling law. With all these powerful methods available, one can envision their application to the engineering of grain boundaries during graphene synthesis. For instance, ideal monocrystalline graphene could be obtained by designing seamless boundaries between coalescing graphene grains. With available large-area monocrystalline graphene, bilayer graphene with controlled stacking order can be constructed by aligned transfer techniques. The relative orientation of the layers can be identified by either low-energy electron diffraction or Raman spectroscopy. This opens a new research direction of bilayer graphene for designing vertical tunneling devices and planar switching devices.

A grain boundary line is a 1D structure consisting of a series of pentagonal, hexagonal, and heptagonal carbon rings. It is possible to selectively functionalize as well as deposit designed materials only at the GGBs due to their higher chemical reactivity compared to ideal basal graphene. This implies that GGBs can be a good template for the synthesis of 1D materials. Atomic layer deposition, whose precursor is quite inert with the graphene basal plane, would be a good method for the synthesis of sub-nanometer 1D metals and semiconductors.

Another research direction to utilize grain boundaries is to control their density to design sensors for detecting gases and molecules under different environmental conditions. As revealed by our numerical simulations and our experimental measurements, the transport properties of grain boundaries can be strongly altered with chemical modifications of the grain boundaries. Together with highly conductive graphene, electro-biochemical sensing devices with high sensitivity and selectivity could be designed.

Membrane science is another open research area. Although the ideal hexagonal graphene lattice impedes the diffusion of gases, defect sites such as heptagons, octagons, vacancies, and divacancies allow selective diffusion of limited gases and molecules, as mentioned above. This provides new opportunities to explore ultrafine membrane performance via the controlled engineering of grain boundaries and point defects.

Although much progress has been made in the visualization and electrical characterization of GGBs from atomic scale to macro scale, issues still remain. The structure of GGBs is determined by the different orientations between merging domains, and the related physical and chemical properties are predicted to be strongly chirality-dependent. However, no electrical measurements have revealed such effects. The question is whether this originates from a device fabrication process which inevitably functionalizes GGBs, or if the native structure of GGBs is disordered, different from theoretical predictions.

GGBs also present challenges for the development of large scale graphene-based spintronic devices,[99] and for harvesting the unique optical properties of graphene. For instance, GGBs introduce non-trivial local symmetry breaking which could significantly impact spin/pseudospin coupling and spin relaxation times, as well as the formation and propagation of plasmonic excitations. Similarly, the peculiar structure of interconnected GGBs could affect transport properties in high magnetic fields, such as the quantum Hall effect. Overall, controlling the atomic structure of GGBs by CVD is a big challenge from a scientific point of view, but would be a huge step forward in the realization of next-generation technologies based on this material.

**Acknowledgements**

The research leading to these results has received funding from the European Union Seventh Framework Programme under grant agreement number 604391 Graphene Flagship. A.W.C. and S.R. acknowledge support from the Spanish Ministry of Economy and Competitiveness under contract MAT2012-33911, and from the SAMSUNG Global Innovation Program. J.K. acknowledges support from the Austrian Science Fund (FWF) through LM1481-N20 and from University of Helsinki funds. D.L.D., V.L.N.and Y.H.L. acknowledge the Institute for Basic Science (EM1304), the HRD Program (No. 20124010203270) of the KETEP grant funded by the Korean Ministry of Knowledge Economy, and BK-Plus through the Ministry of Education, Korea. A. W. Cumings and D. L. Duong contributed equally to this work.

Received: ((will be filled in by the editorial staff))
Revised: ((will be filled in by the editorial staff))
Published online: ((will be filled in by the editorial staff))

**Figure captions**

**Figure 1.** Structure and morphology of GGBs by theory, TEM, and STM/AFM. a, Top panel; 5-7 GGB between two graphene grains with a misorientation angle of 21.8$^o$. Bottom panel; TEM image[23] of a thin 5-7 GGB between grains with a misorientation angle of 27$^o$. Reproduced with permission.[23] Copyright 2011, Nature Publishing Group. b, Left panel; simulated construction of a disordered GGB, including a range of non-hexagonal rings and carbon vacancies.[37] Right panel; STM image of a disordered GGB revealing a similar morphology to the simulated one. Reproduced with permission.[38] Copyright 2012, AIP Publishing. c, Top panel; 3D morphology of a 5-7 GGB, indicating out of plane relaxation.[39] Bottom panels; buckled AFM morphology of polycrystalline graphene after UV exposure. Position 2 indicates out of plane buckling at the GGB.[40] Reproduced with permission.[40] Copyright 2012, Nature Publishing Group. d, The simulated patterns and STM images of two merged grains with identical orientation on a BN substrate (top panels) and a Ni substrate (bottom panels).[64,65] No GGB is observed on the BN substrate, while a 5-8-5 GGB line appears on the Ni substrate. Reproduced with permission.[64,65] Copyright 2013 and 2010, Nature Publishing Group.

**Figure 2.** TEM approach to identifying graphene grain orientations. a, an electron diffraction pattern arising from two misoriented grains. b, Mapping of several grains with different orientations. Reproduced with permission.[24] Copyright 2011, ACS Publishing.

**Figure 3.** Liquid crystal coating approach to identifying graphene grain orientations. a, The hexagonal rings of LC molecules align coherently with hexagonal rings in graphene. Reproduced with permission.[67] Copyright 2012, Nature Publishing Group. b, POM images of LC molecules aligned on each graphene grain, revealing a strong optical contrast between misoriented grains.

**Figure 4.** UV treatment approach to identifying graphene grain orientations. a, Principle of GGB visualization by UV treatment. b-c, Selective oxidation of an underlying the copper substrate for direct optical identification (b) of the GGBs, confirmed by AFM (c). Reproduced with permission.[40] Copyright 2013, Nature Publishing Group.

**Figure 5.** Two-probe measurement of GGBs. a, Differential tunneling conductance at various points on (blue lines) and around (red lines) a GGB. The appearance of defect states is evident on the GGBs. Reproduced with permission.[75] Copyright 2013, Elsevier Publishing. b, STM image of the GGB studied in panel a, where the colored dots indicate the positions of dI/dV measurements. c, dI/dV map across a GGB. d, Location of the dI/dV minimum as a function of tip position, indicating the presence of an electrostatic barrier at the GGB. Reproduced with permission.[74] Copyright 2013, ACS Publishing.

**Figure 6.** Principle of four-probe measurement applied to GGBs. A serie of Hall bars is fabricated across the GGB region. The resistivity of the GGBs can be extracted from this measurement set-up. Intra-grain resistances $R_L$ and $R_R$ are subtracted from the inter-grain resistance to obtain $R_B$, the resistance of the GGB.

**Figure 7**. Four-probe measurement of GGBs. a, Example of a four-probe setup for measuring the resistivity of a GGB. b, I-V curves measured within individual grains (red and blue curves) and across the GGB (green curve). The reduced slope for the inter-grain measurement indicates extra resistance contributed by the GGB. Reproduced with permission.[53] Copyright 2011, Nature Publishing Group. c, Four-probe measurement setup mounted on a TEM holder, where individual graphene grains are identified in the red and blue regions. d, Top plot; four-probe measurements of the inter- and intra-grain resistance as a function of gate voltage (black and gray curves, respectively). Bottom plot; the extracted GB resistivity as a function of gate voltage in volt. Reproduced with permission.[50] Copyright 2013, AAAS.

**Figure 8.** Principle of the scaling law to extract the GGB resistivity. a, Derivation of the ohmic scaling law. b-c, Sheet resistance measurements of graphene with small and large grain sizes. d, Extraction of GGB resistivity by fitting the scaling law to sheet resistance measurements.

**Figure 9.** Global measurements from scaling law. a, Sheet resistance of polycrystalline graphene as a function of average grain size. Grain sizes were determined via Raman spectroscopy. Reproduced with permission.[76] Copyright 2011, IOP Publishing. b, Another example of the scaling behavior of polycrystalline graphene. The dotted line represents a fit to the scaling law described in the main text. Reproduced with permission.[40] Copyright 2012, Nature Publishing Group.

**Figure 10**. Chemical reactivity of GGBs by experiments. a, Representation of selective chemical functionalization of GGBs. b, The location of GGBs can be imaged with AFM after burning them away at high temperature, which highlights their selective oxidation. Reproduced with permission.[69] Copyright 2011, AIP Publishing. c, An optical image of polycrystalline graphene indicates the selective oxidation of an underlying copper substrate below the GGBs. d, Raman spectroscopy indicates the strong oxidation at the GGBs after UV treatment. e-f, Raman mapping indicates strong oxidation of the GGBs (D-band), as well as strain due to the expansion of the oxidized copper substrate below the GGBs (G and G' band shifts). Reproduced with permission.[40] Copyright 2012, Nature Publishing Group.

**Figure 11.** Optical image of the four-probe device across a GGB. a, E-beam lithography resist (PMMA) location at a merging region including a GGB. b-c, A final device with Hall bar geometry at merging region of two graphene domains.

**Figure 12.** $O_2$ Selective functionalization of GGBs by UV treatment under environment. a-b, Effect of annealing at 250 $^{o}$C in 3h. Funtional groups are removed from a GGB. c-d, Effect of UV treatment under $O_2$ environment. The exclusive change of the inter-grain resistance indicates selective functionalization at the GGB. The UV treatment is saturated after 1 minute of UV treatment.

**Figure 13.** Brief explanation of the simulation method for theoretical study. a, Small portion of a polycrystalline graphene sample. b-d, Time evolution of a wave packet within the sample.

**Figure 14.** Simulation of the effect of functional groups at GGBs. a-b, Schematic of GGBs functionalized by H and OH groups, respectively. c, Dependence of the resistivity of GGBs on functional groups with various concentrations. d, Summary of experimental and simulated results for the resistivity of GGBs.

**Figure 1**

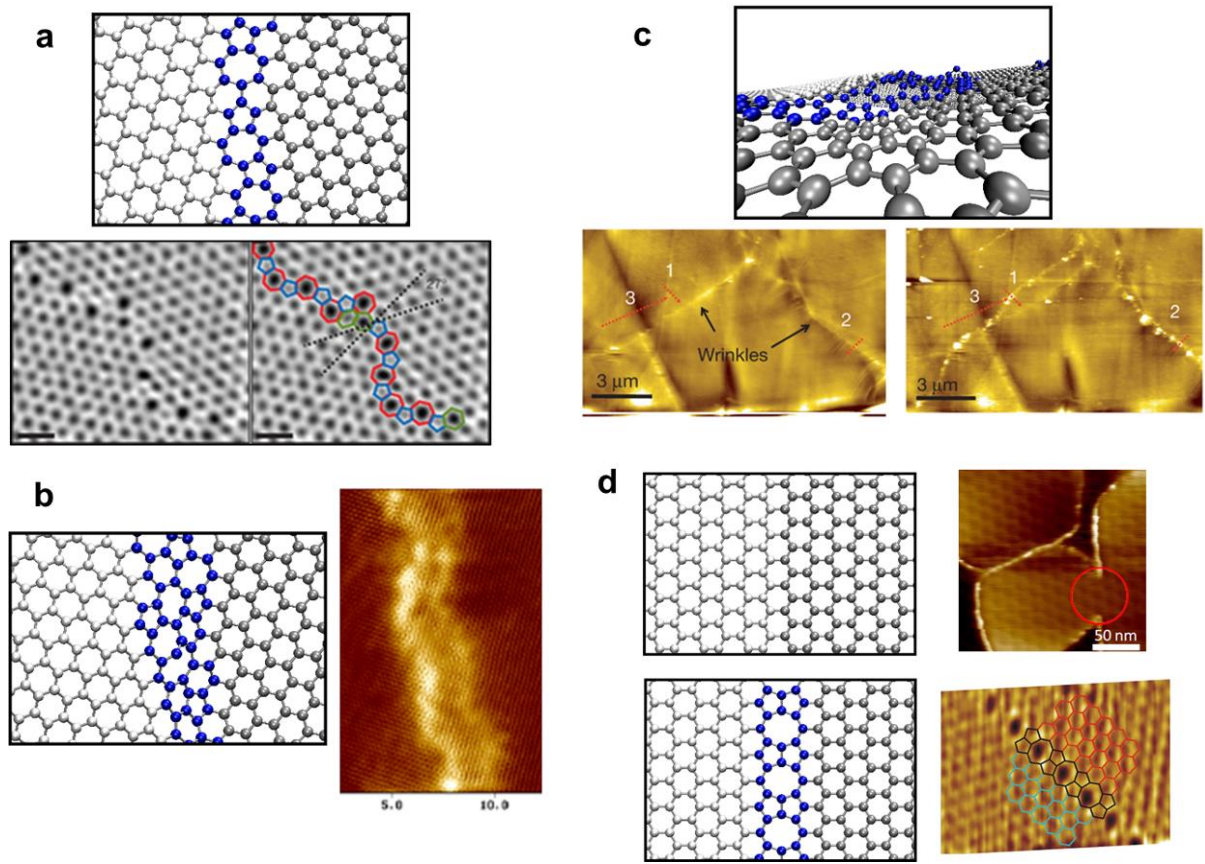

**Figure 2**

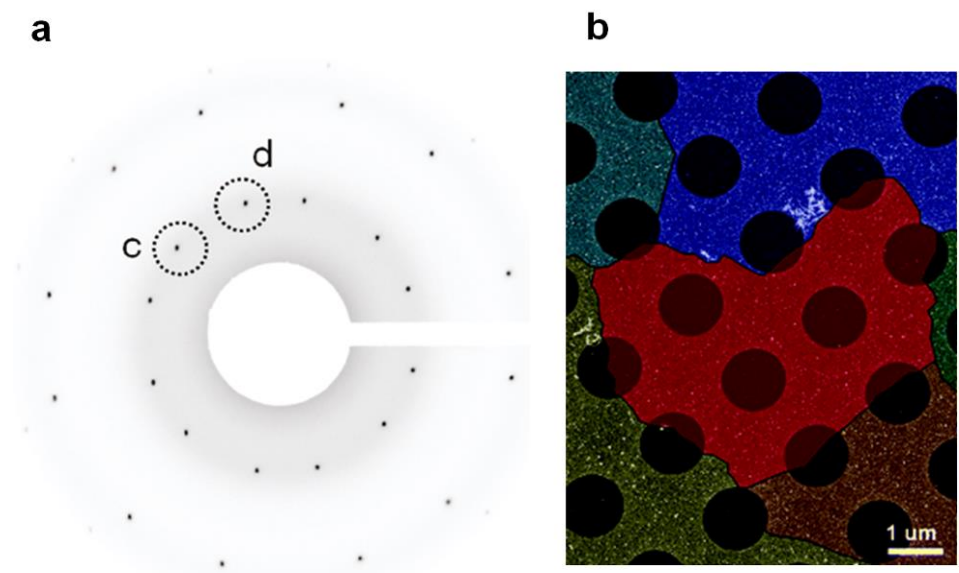

**Figure 3**

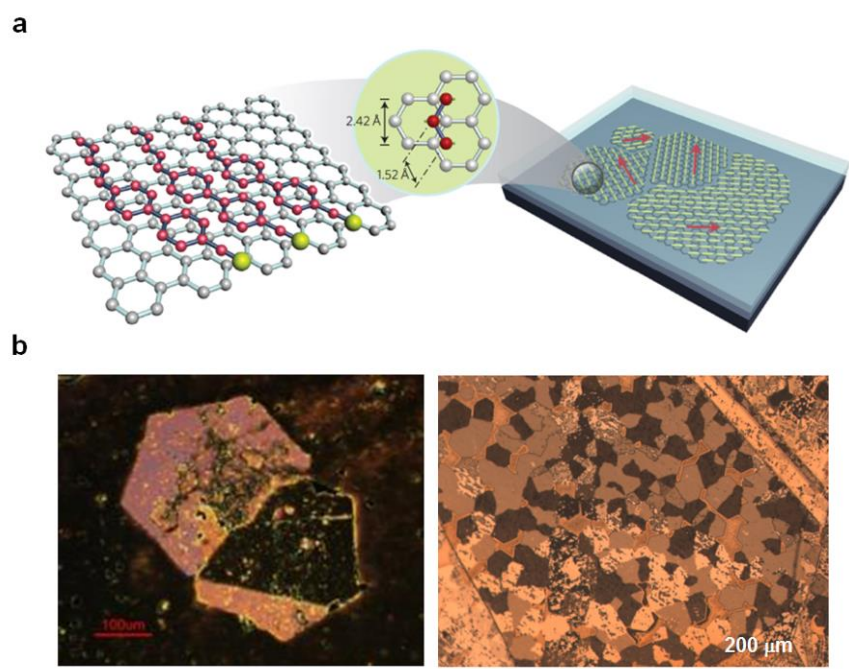

**Figure 4**

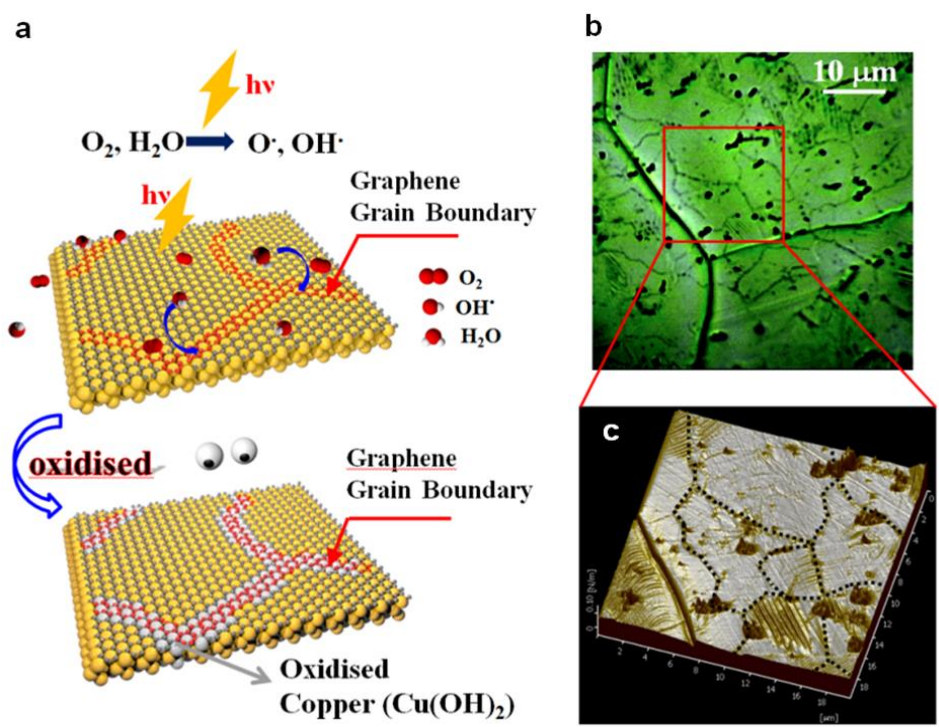

**Figure 5**

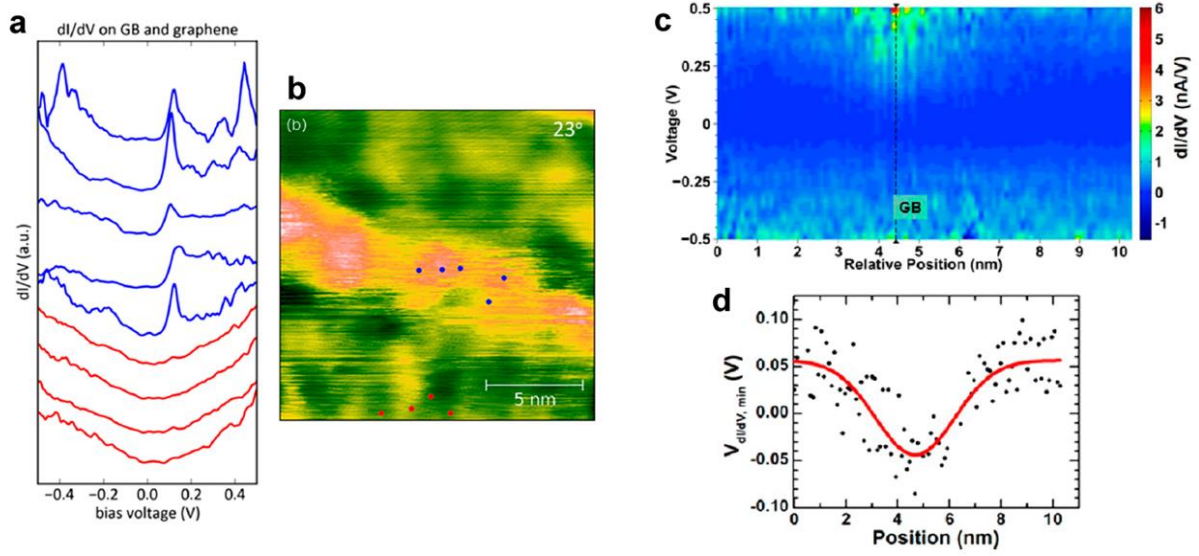

**Figure 6**

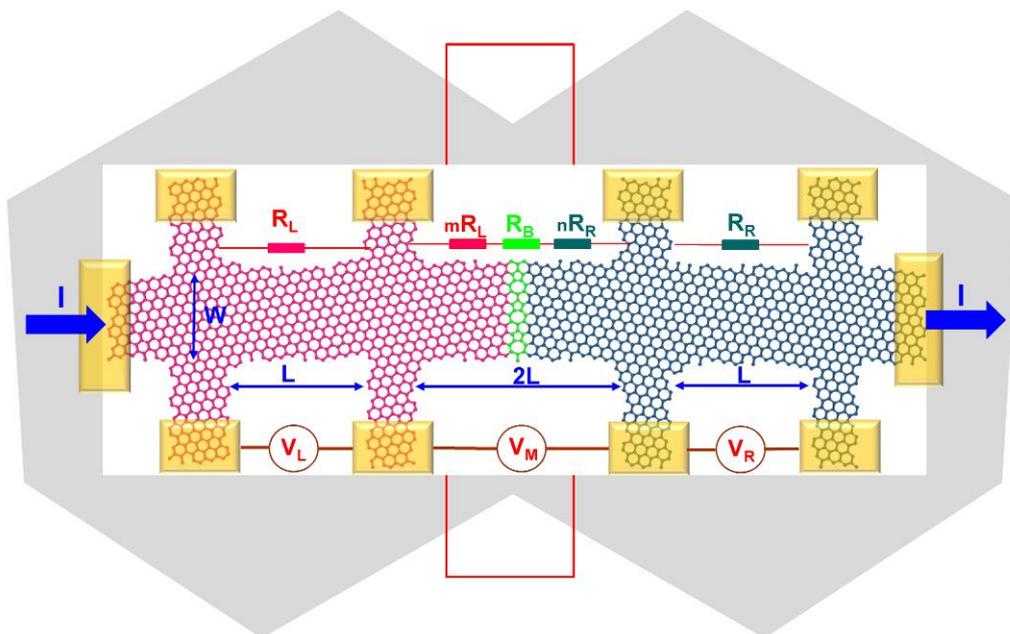

**Figure 7**

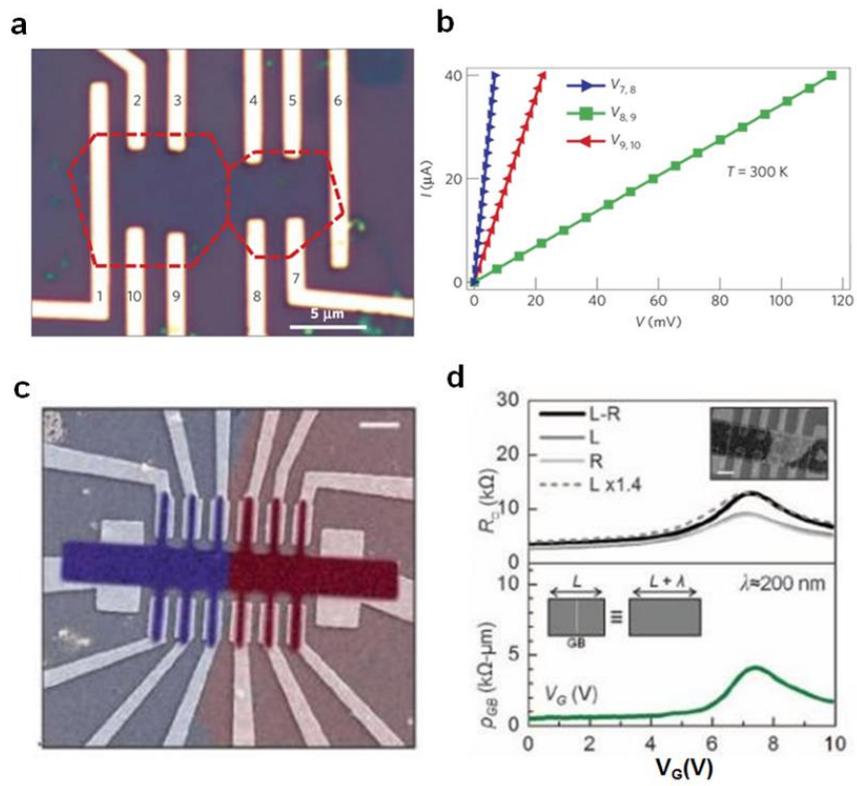

**Figure 8**

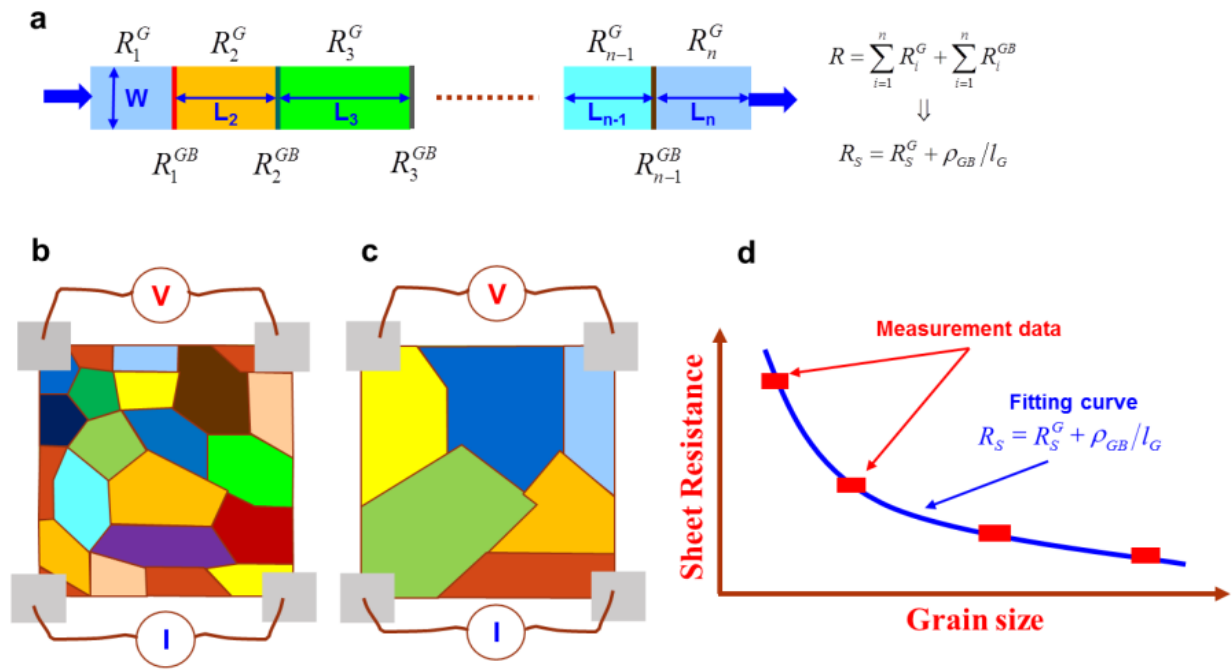

**Figure 9**

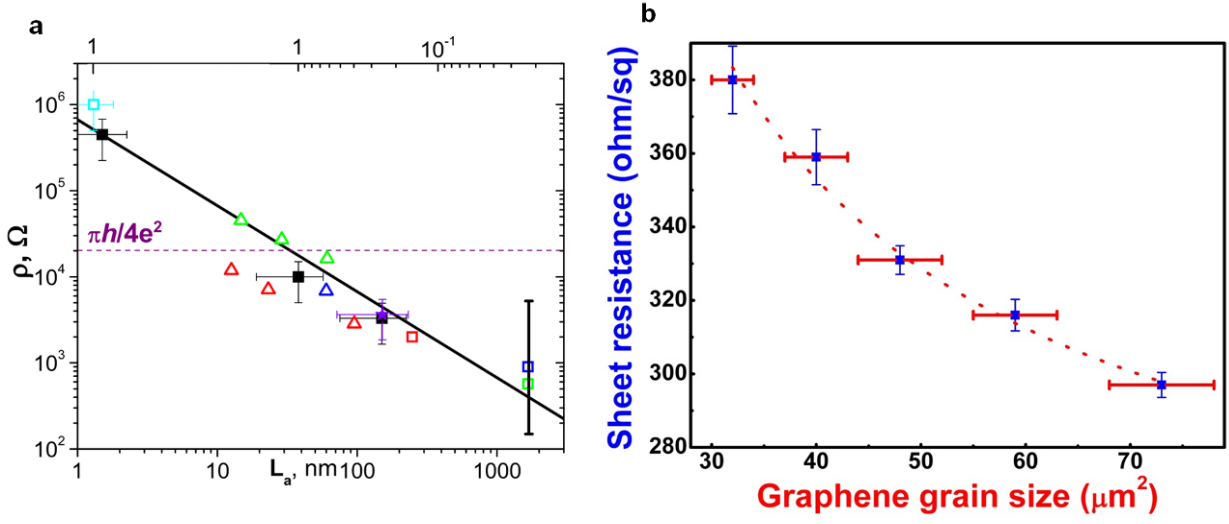

**Figure 10**

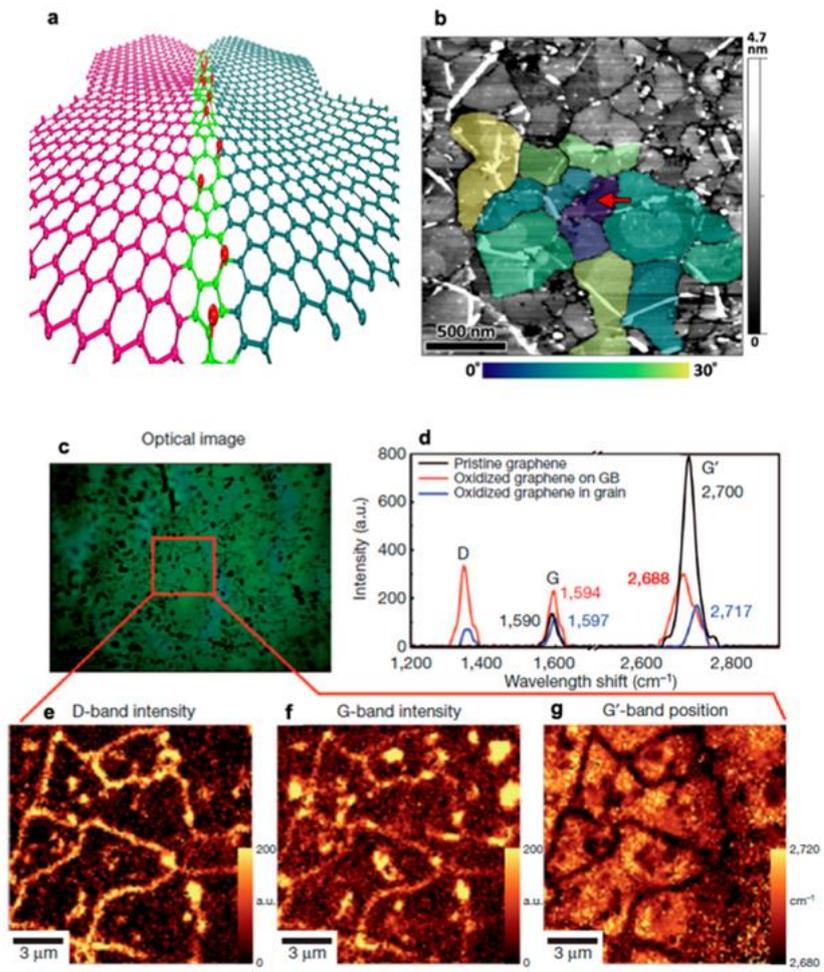

**Figure 11**

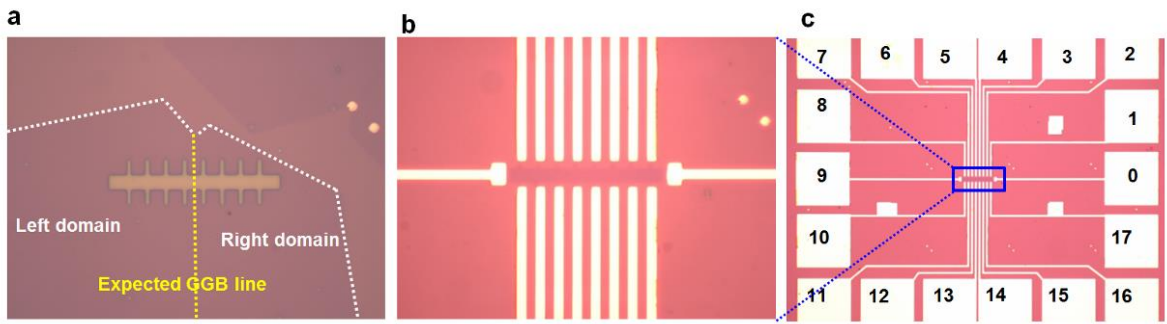

**Figure 12**

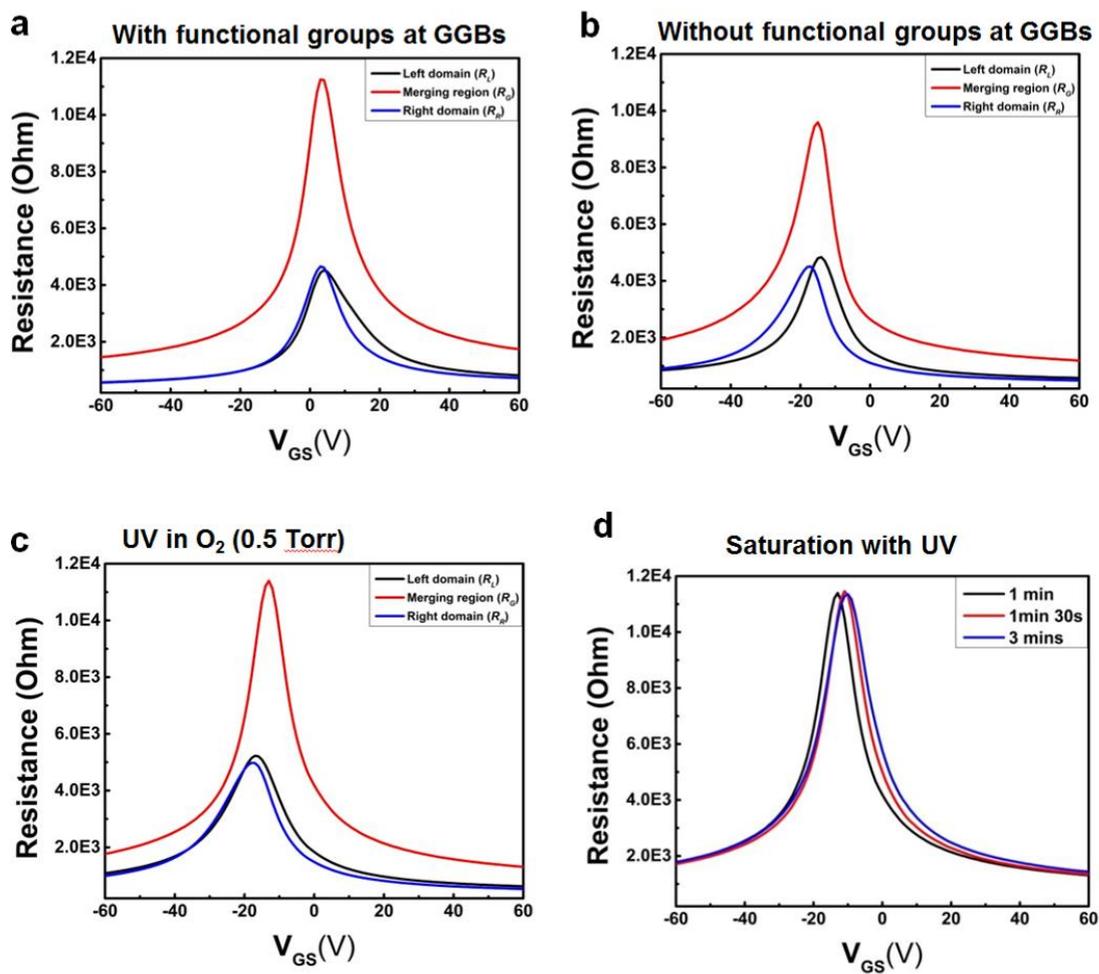

**Figure 13**

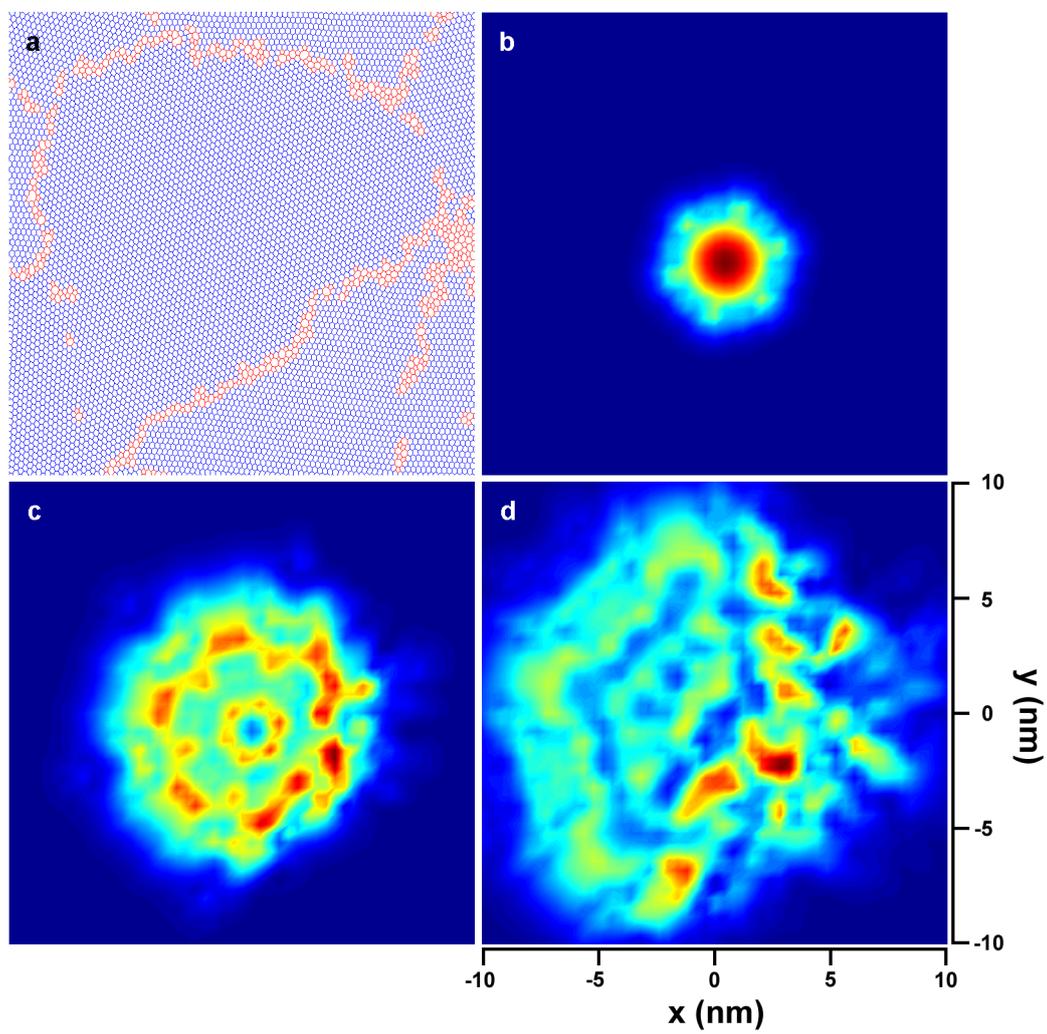

**Figure 14**

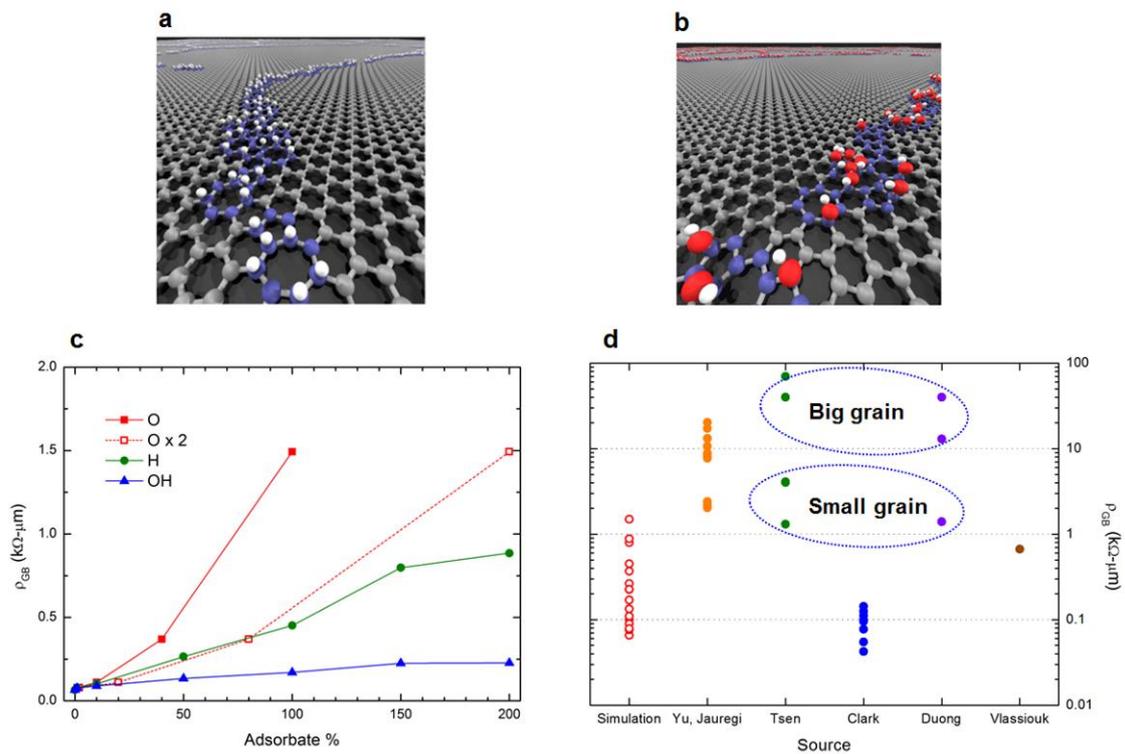

## Authors

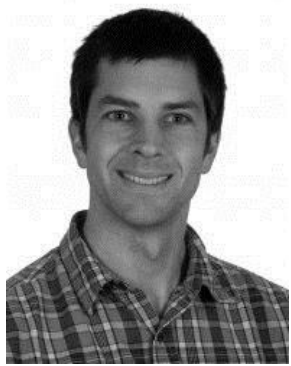

Aron Cummings received his Ph.D. from Arizona State University in 2009 under the supervision of Prof. David Ferry, and his dissertation focused on spin-dependent transport in semiconductor quantum wires. He was a postdoctoral researcher at Sandia National Laboratories from 2010-2013, working with Dr. François Léonard on carbon nanotube nanoelectronic devices. He is now a postdoctoral researcher in Prof. Stephan Roche's group at the Institut Català de Nanociència i Nanotecnologia. His current research focuses on charge and spin transport in disordered graphene.

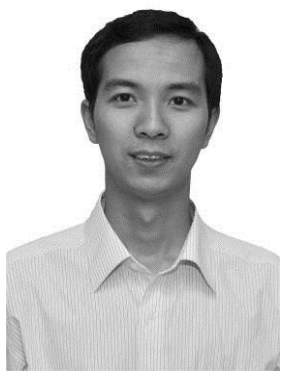

Dinh Loc Duong is a postdoctoral researcher at IBS Center for Integrated Nanostructure Physics (CINAP), Sungkyunkwan University (SKKU) since 2012. He received his Ph.D from Sungkyunkwan Institute of Nanoscience and Nanotechnology (SAINT) in 2012 under the supervision Prof. Young Hee Lee. His current research interests are 2D materials including growth, characterization and applications.

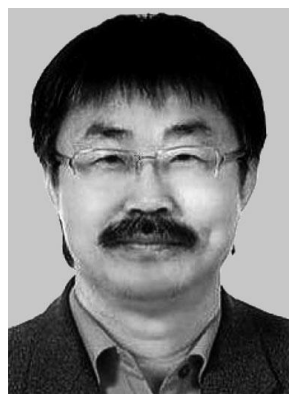

Young Hee Lee is a Professor of Energy Science, Physics department at Sungkyunkwan University. He is the Director of IBS Center for Integrated Nanostructure Physics (CINAP) at Sungkyunkwan University. He received his Ph.D. in physics from Kent State University, USA. His research interests include carrier dynamics, thermoelectric properties, quantum mechanical tunneling properties of CNTs, two dimensional layered materials including graphene, h-BN, transition metal chalcogenides, design of their hybrid structures, and their applications to electronic and optical devices, such as flexible and stretchable transistors and conducting film; energy storage, such as supercapacitors, nanobatteries, and hydrogen storage; and neuroscience and cancer therapy by using noncontact graphene electrical stimulators.

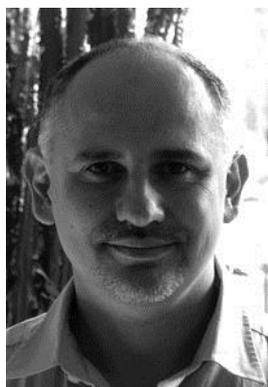

Stephan Roche (1969) is ICREA Research Professor, head of the Theoretical and Computational Nanoscience Group of Catalan Institute of Nanoscience and Nanotechnology (ICN2). He is a theoretical physicist expert in quantum transport and in the development of computational modelling of nanomaterials and devices. His expertise includes the development of order N quantum transport (Kubo and Landauer-Büttiker conductances), with which he has pioneered mesoscopic transport studies in chemically disordered graphene-based materials and devices. He has a deep experience in developing advanced simulation tools in the context of industrial research, with collaborations including large companies such as NEC, STMicroelectronics, and SAMSUNG. He is co-supervising the GRAPHENE SPINTRONICS Workpackage of the Graphene Flagship project.

**The table of contents entry**
**By controlling the structures, distribution and chemical functionalization of grain boundaries of graphene at a nanoscale**, one can envision disruptive technologies and novel fields of research. In this review, we will present fascinating opportunities offered by this versatile material, together with a description of its essential structural and electronic features.

**Keyword** graphene, grain boundaries, charge transport, scaling law, functionalization.

*Aron W. Cummings, Dinh Loc Duong, Van Luan Nguyen, Dinh Van Tuan,*
*Jani Kotakoski, Jose Eduardo Barrios Varga, Young Hee Lee\* and Stephan Roche\**

**Title**
Charge Transport in Polycrystalline Graphene:
Challenges and Opportunities

ToC figure

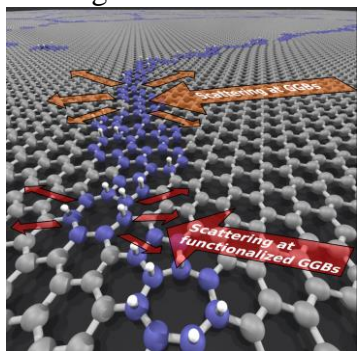